\documentclass[10pt,twocolumn,twoside]{IEEEtran}
\IEEEoverridecommandlockouts
\usepackage{cite}
\usepackage{algorithmic}
\usepackage{textcomp}
\usepackage{xcolor}
\usepackage{soul}
\usepackage{graphicx,color}
\usepackage{amsfonts,amsthm,amssymb}
\usepackage{mathrsfs,amsmath,arydshln,latexsym}
\usepackage{cite,url}
\usepackage[bookmarks,colorlinks]{hyperref}
\usepackage[linesnumbered,ruled,vlined]{algorithm2e}
\usepackage{multirow,array,makecell}
\usepackage{stfloats}
\usepackage{fancyhdr}
\usepackage{bm}
\usepackage{enumitem}
\usepackage{pifont}
\usepackage{subcaption}
\usepackage{nomencl}
\usepackage{colortbl}
\usepackage{diagbox}
\usepackage{balance}
\usepackage{braket}
\usepackage{physics}
\usepackage{framed}
\usepackage{threeparttablex}
\allowdisplaybreaks[4]

\newtheorem{theorem}{Theorem}

\begin{document}

	\title{
		Rydberg Atomic Quantum Receivers for\\ Multi-Target DOA Estimation
	}

	\author{Tierui Gong,~\IEEEmembership{Member,~IEEE}, 
		Chau Yuen,~\IEEEmembership{Fellow,~IEEE},
		Chong Meng Samson See, \\
		Mérouane Debbah,~\IEEEmembership{Fellow,~IEEE},
		Lajos Hanzo,~\IEEEmembership{Life Fellow,~IEEE}
		\vspace{-1.3cm}
		\thanks{T. Gong and C. Yuen are with School of Electrical and Electronics Engineering, Nanyang Technological University, Singapore 639798 (e-mail: trgTerry1113@gmail.com, chau.yuen@ntu.edu.sg). C. M. S. See is with DSO National Laboratories, Singapore 118225 (e-mail: schongme@dso.org.sg). M. Debbah is with KU 6G Research Center, Department of Computer and Information Engineering, Khalifa University, Abu Dhabi 127788, UAE (e-mail: merouane.debbah@ku.ac.ae) and also with CentraleSupelec, University Paris-Saclay, 91192 Gif-sur-Yvette, France. L. Hanzo is with School of Electronics and Computer Science, University of Southampton, SO17 1BJ Southampton, U.K. (e-mail: lh@ecs.soton.ac.uk). 
		}
	}

	\markboth{Accepted by IEEE Transactions on Vehicular Technology, 2025}%
	{Gong \MakeLowercase{\textit{et al.}}: Rydberg Atomic Quantum Receivers for Multi-Target DOA Estimation}

	\maketitle

	\begin{abstract}
		Quantum sensing technologies have experienced rapid progresses since entering the `second quantum revolution'. Among various candidates,  schemes relying on Rydberg atoms exhibit compelling advantages for detecting radio frequency signals. Based on this, Rydberg atomic quantum receivers (RAQRs) have emerged as a promising solution to classical wireless communication and sensing. To harness the advantages and exploit the potential of RAQRs in wireless sensing, we investigate the realization of the direction of arrival (DOA) estimation by RAQRs. Specifically, we first conceive a Rydberg atomic quantum uniform linear array (RAQ-ULA) aided wireless receiver for multi-target DOA detection and propose the corresponding signal model of this sensing system. Our model reveals that the presence of the radio-frequency local oscillator in the RAQ-ULA creates sensor gain mismatches, which degrade the DOA estimation significantly by employing the classical Estimation of Signal Parameters via Rotational Invariant Techniques (ESPRIT). To solve this sensor gain mismatch problem, we propose the Rydberg atomic quantum ESPRIT (RAQ-ESPRIT) relying on our model. 
		Lastly, we characterize our scheme through numerical simulations, where the results exhibit that it is capable of reducing the estimation error of its classical counterpart on the order of $> 400$-fold and $> 9000$-fold in the PSL and SQL, respectively. 
	\end{abstract}


	\vspace{-0.3cm}
	\section{Introduction}
	
	Rydberg atomic quantum receivers (RAQRs) \cite{schlossberger2024rydberg, zhang2024rydberg, gong2024RAQRs} have recently emerged as a new concept for facilitating the wireless communication and sensing by harnessing the unique quantum mechanical properties of Rydberg atoms in detecting the electric fields of radio-frequency (RF) signals. Specifically, a Rydberg atom represents an excited atom having one or more electrons transiting from the ground-state energy level to a higher Rydberg-state energy level. Exploiting these Rydberg atoms, the amplitude, phase, polarization, and even orbital angular momentum of the RF signals have been experimentally captured by RAQRs at an unprecedented precision. Particularly, RAQRs have the potential to revolutionize existing antenna-based RF receivers, such as multiple-antenna systems \cite{Larsson2014Massive,Gong2023Holographic,LiuIntegrated2022}, paving the way for facilitating classical wireless communication and sensing through a quantum-domain solution. 
	
	As a novel quantum solution for wireless systems, RAQRs exhibit numerous fascinating characteristics, including but not limited to super-high sensitivity, extremely-wideband tunability, international system of units (SI) traceability, simultaneous full-circle angular direction detectability, and relatively compact form factor. 
	Specifically, the super-high sensitivity is realized by harnessing the extremely-large dipole moments of Rydberg atoms, which has been experimentally shown to be on the order of nV/cm/$\sqrt{\text{Hz}}$ \cite{jing2020atomic,borowka2024continuous}, outperforming conventional antennas. Next, the extremely-wideband tunability, spanning from direct-current to Terahertz, is enabled by exploiting the electron transitions among the different Rydberg-state energy levels. Additionally, the SI traceability implies that the measurements carried out by RAQRs are directly linked to the SI constants without requiring any calibration. Furthermore, the simultaneous full-circle angular direction detectability reveals that RAQRs are capable of receiving RF signals through a single vapour cell without any angular direction restriction. Lastly, the size of the vapour cell of RAQRs is independent of the RF wavelength. Additionally, implementing more complex receivers, such as, multiple-antenna and multiband schemes, is feasible using a single vapour cell. All these aspects facilitate a compact form factor for RAQRs.

	To exploit the huge potential of RAQRs, experimental studies were carried out for verifying the capabilities of RAQRs. However, the application of RAQRs to wireless sensing, such as direction-of-arrival (DOA) estimation \cite{Huang2018Deep,Shaikh2023DOA}, is not well documented from a signal processing perspective. In \cite{robinson2021determining,Mao2024Digital,Richardson2023}, the initial experimental verifications of RAQRs harnessed for DOA estimation were carried out, demonstrating their feasibility. However, these studies only focus on the experimental design and verification, but lack a general signal model for further guiding the system design and signal processing. Hence, in this article, we unveil the potential of RAQRs for DOA estimation from a signal processing perspective. Specifically, we consider a multi-target scenario and construct a signal model for a RAQR based uniform linear array (RAQ-ULA) aided system. The proposed model serves as a general basis for designing various DOA estimation algorithms. More particularly, our model reveals that the RF local oscillator (LO) of the RAQ-ULA imposes sensor gain mismatches, which cannot be adequately addressed by the classical Estimation of Signal Parameters via Rotational Invariant Techniques (ESPRIT). Therefore, we propose a Rydberg atomic quantum ESPRIT (RAQ-ESPRIT) for solving this problem. Finally, we perform rich numerical simulations to demonstrate the superiority of RAQRs compared to their conventional counterparts.

	\textit{Organization and Notations}: 
	In Section \ref{sec:SigsModel}, we propose the signal model of a sensor array formed by superheterodyne RAQRs. In Section \ref{sec:DOA}, we propose the RAQ-ESPRIT method for DOA estimation. We then present our simulation results in Section \ref{sec:Simulations}, and finally conclude in Section \ref{sec:Conclusions}. 
	The notations are: $\mathscr{R} \{ \cdot \}$ and $\mathscr{I} \{ \cdot \}$ represent the real and imaginary parts of a complex number; $\chi'$ represents the derivative of $\chi$; $\hbar$ denotes the reduced Planck constant and $\jmath^2 = 1$; $c$, $\epsilon_0$ and $Z_0$ are the speed of light in free space, the vacuum permittivity, and the free-space impedance, respectively; $\Tr (\cdot)$ is the trace operator for matrices and $\odot$ represents the Hadamard product.

	\begin{figure}[t!]
		\centering
		\includegraphics[width=0.43\textwidth]{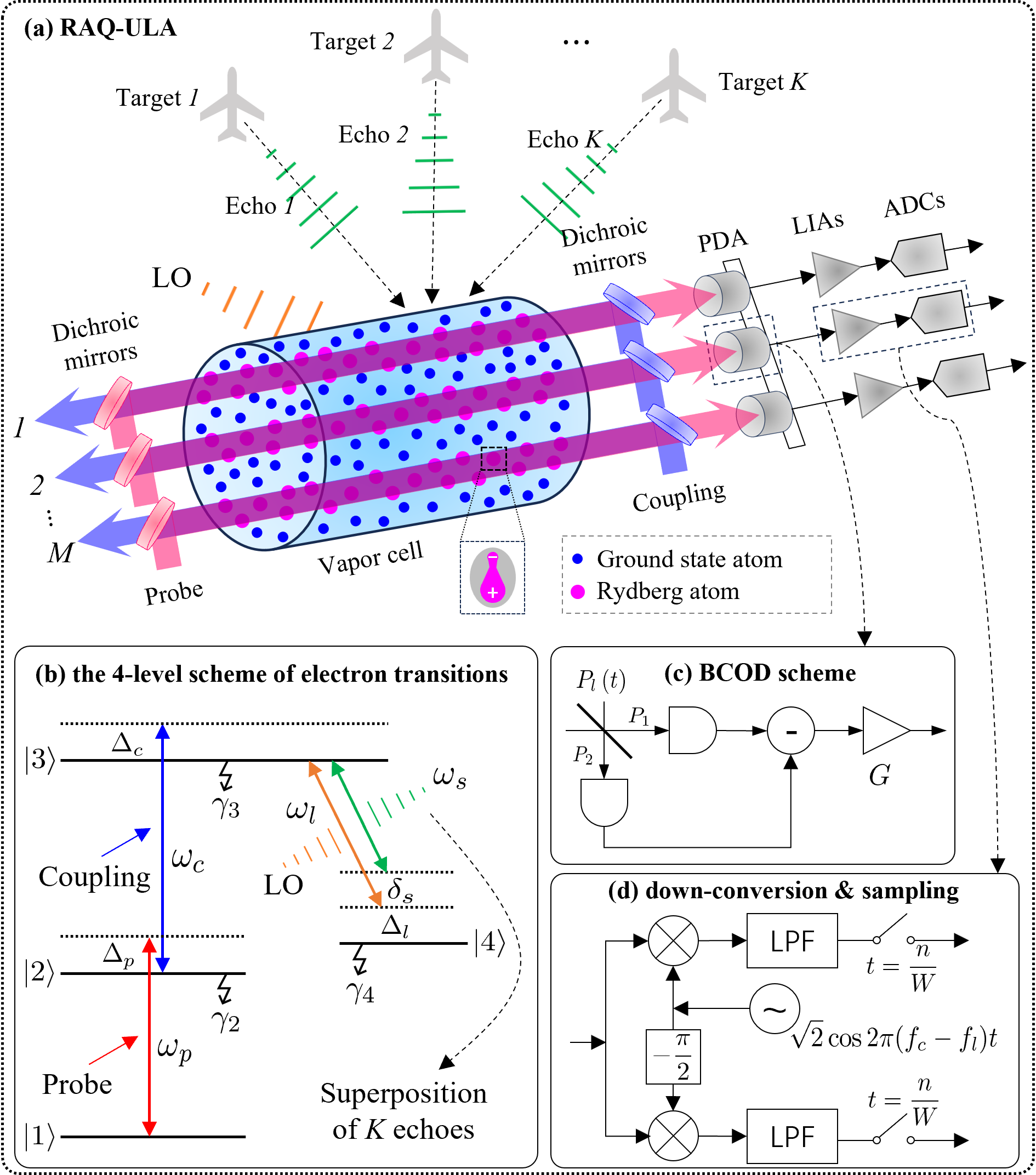}
		\caption{(a) The RAQ-ULA, (b) four-level electron transitions, (c) BCOD scheme, and (d) the down-conversion \& sampling.}
		\vspace{-1.5em}
		\label{fig:RAQMIMOScheme}
	\end{figure}

	\vspace{-0.3em}
	\section{Signal Model of RAQ-ULA Systems} 
	\label{sec:SigsModel}

	We apply the superheterodyne philosophy for our RAQ-ULA system benefiting from  its super-high sensitivity and capability in capturing both the amplitude and phase of RF signals. The structure of the RAQ-ULA is highlighted in Fig \ref{fig:RAQMIMOScheme}(a). Therein, the probe and coupling laser beams are split into $M$ branches, respectively, to form $M$ beam pairs. Each beam pair counter-propagates through the vapour cell to form a receiver sensor that encompasses a multitude of well-prepared Rydberg atoms. We assume that the distance between adjacent receiver sensors is $d$. 
	Furthermore, we consider a total of $K$ targets that produce $K$ echoes to the RAQ-ULA. The echoes and the LO simultaneously influence the Rydberg atoms, which affect the probe beams to be detected by a photodetector array (PDA). Both the amplitude and phase of the echoes are embedded into the detected probe beam, which are extracted by a parallel bank of lock-in amplifiers (LIAs).

	\vspace{-0.3cm}
	\subsection{Quantum Response of Rydberg Atoms} 
	\label{subsec:QR}
	
	The quantum response of each Rydberg atom can be described by a four-level scheme, as shown in Fig. \ref{fig:RAQMIMOScheme}(b). Briefly, the ground state $\ket{1}$, exited state $\ket{2}$, and the pair of Rydberg states $\ket{3}$, $\ket{4}$ are coupled by the probe beam, the coupling beam, and by the superimposed RF signal (echoes + LO), respectively. Specifically, the probe beam has a Rabi frequency of $\Omega_{p}$ and a frequency detuning of $\Delta_{p}$, where the former is directly related to the amplitude of the probe beam, while the latter represents a small frequency shift compared to the $\ket{1} \rightarrow \ket{2}$ transition frequency. Particularly, the probe beam is perfectly resonant with the $\ket{1} \rightarrow \ket{2}$ transition when $\Delta_{p} = 0$. We assume that $\{ \Omega_{p}, \Delta_{p} \}$ are identical for the Rydberg atoms in all the $M$ receiver sensors, so that we can neglect the index $m$. Likewise, we define $\{ \Omega_{c}, \Delta_{c} \}$ for the coupling beam, and $\{ \Omega_{l}, \Delta_{l} \}$ for the LO, respectively. We emphasize that a plane-wave propagation of the LO signal is considered\footnote{We note that this is possible by employing the scheme, where the LO is imposed using a parallel-plate \cite{simons2019embedding}.}, so that all $M$ sensors have the same $\Omega_{l}$. 
	
	Additionally, we consider the plan-wave propagation of the echoes, so that all $M$ sensors experience the same amplitude for each echo. Furthermore, we assume that the LO has a much higher intensity than the echoes, so that the weak echoes having Rabi frequencies of $\Omega_{k}$, $k=1, \cdots, K$ yield a coupling $\sum_{k=1}^{K} \Omega_k \cos{(2 \pi f_{\delta} t + \theta_{\delta, k,m})}$, where $\theta_{\delta,k,m} = \theta_{k} - \theta_{l,m}$ and $f_{\delta} = f_c - f_l$ represent the phase shift difference and frequency difference between the $k$-th echo and the LO impinging on the $m$-th sensor. We note that $\theta_{k}$, $\theta_{l,m}$, $f_c$, and $f_l$ represent the phase of the $k$-th target echo, the phase of the LO at the $m$-th sensor, the carrier frequency of the target echoes, and the carrier frequency of the LO, respectively. 
	Therefore, we express the Rabi frequency of the superimposed RF signal as $\Omega_{\text{RF}, m} \approx \Omega_{l} + \sum_{k=1}^{K} \Omega_k \cos{(2 \pi f_{\delta} t + \theta_{\delta, k,m})}$\footnote{This can be roughly derived based on \cite{gong2024RAQRModel_Journal} by replacing the Rabi frequency related term of a single RF signal with those of multi-target signals. A similar derivation process for the multi-user uplink can be found in \cite{gong2025RAQ_MIMO_Journal}.}. This approximation is facilitated by $\Omega_{l} \gg \sum_{k=1}^{K} \Omega_{k}$. 
	Let us denote the spontaneous decay rate of the $i$-th level by $\gamma_{i}$, $i = 2, 3, 4$, the relaxation rates related to the atomic transition effect and collision effect by $\gamma$ and $\gamma_c$, respectively. For simplicity, we assume $\gamma = \gamma_c = 0$, and $\gamma_3 = \gamma_4 = 0$ as they are comparatively small and hence can be reasonably ignored.

	The excitation and decay of a Rydberg atom will finally reach a balance, where the steady-state solution of the probability density can be characterized by solving the Lindblad master equation (LME). 
	The steady-state solution of the LME for Rydberg atomic quantum single-input single-output (RAQ-SISO) systems has been derived in \cite{gong2024RAQRModel_Journal}. Following a similar derivation process, we can derive the steady-state solution for our RAQ-ULA system. Specifically, we can reuse the result of the susceptibility derived in \cite{gong2024RAQRModel_Journal} for the $m$-th sensor. Explicitly, we have the expression presented as follows 
	\begin{align}
		\nonumber
		\chi_{m} (\Omega_{\text{RF},m}) 
		&= \varsigma \left[ \frac{{{A_{1,m}}\Omega _{{\rm{RF}},m}^4 + {A_{2,m}}\Omega _{{\rm{RF}},m}^2 + {A_{3,m}}}}{{{C_{1,m}}\Omega _{{\rm{RF}},m}^4 + {C_{2,m}}\Omega _{{\rm{RF}},m}^2 + {C_{3,m}}}} \right. \\
		\label{eq:Susceptibility}
		&\left. \quad - \jmath \frac{{{B_{1,m}}\Omega _{{\rm{RF}},m}^4 + {B_{2,m}}\Omega _{{\rm{RF}},m}^2 + {B_{3,m}}}}{{{C_{1,m}}\Omega _{{\rm{RF}},m}^4 + {C_{2,m}}\Omega _{{\rm{RF}},m}^2 + {C_{3,m}}}} \right]. 
	\end{align}
	In \eqref{eq:Susceptibility}, we have $\varsigma = - \frac{2 N_0 \mu_{12}^{2}}{\epsilon_0 \hbar}$, where $N_0$ is the atomic density in the vapour cell and $\mu_{12}$ is the dipole moment of the transition $\ket{1} \rightarrow \ket{2}$. $A_{1,m}$, $A_{2,m}$, $A_{3,m}$, $B_{1,m}$, $B_{2,m}$, $B_{3,m}$, $C_{1,m}$, $C_{2,m}$, $C_{3,m}$ are coefficients obtained so that $\chi_{m} (\Omega_{\text{RF},m})$ can be expressed as a function of $\Omega_{\text{RF},m}$. These coefficients are related to the laser's/LO's configuration parameters (Rabi frequencies and detuning frequencies) and the atomic decay rate, which are detailed in the Appendix A of \cite{gong2024RAQRModel_Journal}.

	\subsection{RF-to-Optical Transformation Model of RAQ-ULA}

	The atomic vapour serves as an RF-to-optical transformer. Specifically, the RF signal perturbs the Rydberg states through strong dipole interactions, thereby modifying the associated atomic coherences and populations. These changes directly influence the atomic susceptibility, which supports the conversion of microscopic dipole responses into the macroscopic probe-beam modulation. Consequently, the RF-induced variations of susceptibility manifest themselves as amplitude attenuation via its imaginary component and phase shifts via its real component, hence facilitating the RF-dependent control of the probe beam. 
	Let us denote the amplitude, frequency, and phase of the $m$-th probe beam at the input of the vapour cell by $\{ U_{0,m}, f_{p}, \phi_{0,m} \}$, respectively. Their output counterparts are given by \cite{gong2024RAQRModel_Journal} 
	\begin{align}
		\label{eq:AmplitudeRelation}
		U_{p, m} (\Omega_{\text{RF},m}) 
		&= U_{0,m} e^{ - \frac{\pi \ell}{\lambda_p} \mathscr{I} \{\chi_{m} (\Omega_{\text{RF},m}) \} }, \\
		\label{eq:PhaseRelation}
		\phi_{p, m} (\Omega_{\text{RF},m})  
		&= \phi_{0,m} + \tfrac{\pi \ell}{\lambda_p} \mathscr{R} \{\chi_{m} (\Omega_{\text{RF},m}) \}, 
	\end{align}
	where $\lambda_p$ is the wavelength of the probe beam, $\ell$ is the length of the vapor cell. \eqref{eq:AmplitudeRelation} is known as the Lambert-Beer law.

	Given the amplitude and phase of the output probe beam at the $m$-th sensor in \eqref{eq:AmplitudeRelation} and \eqref{eq:PhaseRelation}, we formulate its waveform as  
	\begin{align}
		\nonumber
		P_{m} (\Omega_{\text{RF},m}, t) 
		&= \sqrt{2\mathcal{P}_{m} (\Omega_{\text{RF},m})} \cos \left( 2 \pi f_{p} t + \phi_{p,m} (\Omega_{\text{RF},m}) \right) \\
		\label{eq:OutputProbeSig}
		&= \sqrt{2} \mathscr{R} \left\{ P_{b,m} (\Omega_{\text{RF},m}, t) e^{\jmath 2 \pi f_p t} \right\},
	\end{align}
	where $\mathcal{P}_{m} (\Omega_{\text{RF},m}) = \frac{\pi c \epsilon_0}{8 \ln {2}}  F_p^2 \left| U_{p,m}(\Omega_{\text{RF},m}) \right|^{2}$ represents the power of the output probe beam, $F_{p}$ is the full width at half maximum (FWHM) of the probe beam. Additionally, $P_{b,m} (\Omega_{\text{RF}}, t) \triangleq \sqrt{\mathcal{P}_{m} (\Omega_{\text{RF},m})} e^{\jmath \phi_p (\Omega_{\text{RF},m})}$ represents the equivalent baseband signal of the output probe beam. 
	
	All $M$ output probe beams are further detected by the PDA, where each element of the PDA obeys a balanced coherent optical detection (BCOD) scheme \cite{gong2024RAQRModel_Journal}. In each PDA element, the probe beam is mixed with a local optical beam ${P_{m}^{(l)}}\left( t \right)$ to form two distinct optical signals, ${P_1} = \frac{1}{{\sqrt 2 }}[ {P_{m}^{(l)}}\left( t \right) - P_{m} (\Omega_{\text{RF},m}, t) ]$ and ${P_2} = \frac{1}{{\sqrt 2 }}[ {P_{m}^{(l)}}\left( t \right) + P_{m} (\Omega_{\text{RF},m}, t) ]$, which are then detected by two photodetectors, respectively. The consequent photocurrents are subtracted and amplified by a low-noise amplifier (LNA) having a gain of $G$, as shown in Fig. \ref{fig:RAQMIMOScheme}(c). The $M$ PDA outputs are then down-converted to baseband signals through the LIAs in a point-to-point manner, where the intermediate frequency $f_c - f_l$ is removed. Finally, the analog baseband signals are sampled to obtain its discrete counterpart. Both the down-conversion and sampling are shown in Fig. \ref{fig:RAQMIMOScheme}(d).

	\vspace{-0.3cm}
	\subsection{The Proposed Signal Model of RAQ-ULA}
	
	Based on the signal model of RAQ-SISO systems constructed in \cite{gong2024RAQRModel_Journal}, we know that the $k$-th received target echo will be affected by a gain and a phase shift of the $m$-th receiver sensor. \textit{The gain and phase shift are jointly determined by both the atomic response and the specific photodetection scheme selected}. As we employ the BCOD scheme, we obtain the gain and phase shift of the $m$-th receiver sensor as formulated 
	\begin{align}
		\label{eq:GainBCOD}
		\varrho_{m} &= 4 \alpha_{1}^{2} Z_0 G \mathcal{P}_{m}^{(l)} \mathcal{P}_{m} (\Omega_{l}) {\kappa_{m}^{2}}({\Omega _l}), \\
		\label{eq:PhaseBCOD}
		\Phi_{m} &= \frac{ e^{ - \jmath \left[ {\theta_{l,m}} - \varphi_{m} ({\Omega _l}) \right] } }{2} + \frac{ e^{ - \jmath \left[ {\theta_{l,m}} + \varphi_{m} ({\Omega _l}) \right] } }{2}, 
	\end{align}
	where $\varrho_{m}$ is jointly determined by the LNA gain $G$, the power of the local optical beam $\mathcal{P}_{m}^{(l)}$, the power of the output probe beam $\mathcal{P}_{m} (\Omega_{l})$, and the atomic responsivity ${\kappa_{m}}({\Omega _l})$;  $\Phi_{m}$ is jointly determined by the phase of the LO signal ${\theta_{l,m}}$, and the superimposed phase $\varphi_{m} ({\Omega _l})$ between the local optical beam and the phase of the probe beam influenced by the atomic vapor. Furthermore, we have $\alpha_{1} \triangleq \frac{{\eta q}}{{\hbar \omega_p }}$ and 
	\begin{align}
		\label{eq:Kappa2}
		\kappa_{m} (\Omega_l) 
		&= \alpha_{2} \sqrt {{{\left[ {{\mathscr R}\{ \chi'_{m} ({\Omega _l})\} } \right]}^2} + {{\left[ {{\mathscr I}\{ \chi'_{m} ({\Omega _l})\} } \right]}^2}}, \\
		\label{eq:varphi_2}
		{\varphi_{m} ({\Omega _l})} 
		&= {\phi_{m}^{(l)}} - {\phi_{p, m}}({\Omega _l}) + {\psi_{p, m}}({\Omega _l}). 
	\end{align}
	In \eqref{eq:Kappa2} and \eqref{eq:varphi_2}, we have $\alpha_{2} \triangleq \frac{\pi \ell \mu_{34}}{\hbar \lambda_p}$, ${\phi_{m}^{(l)}}$ denotes the phase of the local optical beam in the $m$-th channel, and 
	\begin{align}
		\nonumber
		&\mathscr{R} \{ \chi'_{m} \left( \Omega_{l} \right) \} 
		= 2 \varsigma  {\Omega_{l}} \left[ \frac{ 2{A_{1,m}}\Omega_{l}^2 + {A_{2,m}} }{{{C_{1,m}}\Omega _{l}^4 + {C_{2,m}}\Omega _{l}^2 + {C_{3,m}}}} \right. \\
		\nonumber
		&\left. - \frac{{ \left( {{A_{1,m}}\Omega _{l}^4 + {A_{2,m}}\Omega _{l}^2 + {A_{3,m}}} \right) \left( {2{C_{1,m}}\Omega_{l}^2 + {C_{2,m}}} \right)}}{{{{\left( {{C_{1,m}}\Omega_{l}^4 + {C_{2,m}}\Omega _{l}^2 + {C_{3,m}}} \right)}^2}}} \right], \\
		\nonumber
		&\mathscr{I} \{ {\chi'_{m}} \left( \Omega_{l} \right) \}
		= - 2 \varsigma {\Omega_{l}} \left[ \frac{ 2{B_{1,m}}\Omega_{l}^2 + {B_{2,m}} }{{{C_{1,m}}\Omega_{l}^4 + {C_{2,m}}\Omega _{l}^2 + {C_{3,m}}}} \right. \\
		\nonumber
		&\left. - \frac{{ \left( {{B_{1,m}}\Omega _{l}^4 + {B_{2,m}}\Omega _{l}^2 + {B_{3,m}}} \right) \left( {2{C_{1,m}}\Omega_{l}^2 + {C_{2,m}}} \right)}}{{{{\left( {{C_{1,m}}\Omega _{l}^4 + {C_{2,m}}\Omega _{l}^2 + {C_{3,m}}} \right)}^2}}} \right], \\
		\nonumber
		&\psi_{p, m} (\Omega_{l}) = \arccos \frac{ {{\mathscr I}\{ \chi'_{m} ({\Omega _l})\} } }{ \sqrt {{{\left[ {{\mathscr I}\{ \chi'_{m} ({\Omega _l})\} } \right]}^2} + {{\left[ {{\mathscr R}\{ \chi'_{m} ({\Omega _l})\} } \right]}^2}} }.
	\end{align}

	Based on the above discussion, we know that all $K$ target echoes will be affected by the gain \eqref{eq:GainBCOD} and phase shift \eqref{eq:PhaseBCOD} at the $m$-th sensor. Let us first select the receiver sensor indexed by $m = 1$ as the reference sensor. Then, we can formulate the signal model of the $m$-th receiver sensor as follows 
	\begin{align}
		\label{eq:NarrowbandSampledOutputPlusNoise}
		y_{m} = \sqrt{\varrho_m} \Phi_m \sum_{k=1}^{K} A_{m,k} \left( \theta_{k} \right) s_{k} + w_m, 
	\end{align}
	where $A_{m,k} \left( \theta_{k} \right) = \exp( \jmath \frac{2 \pi}{\lambda} (m-1) d \sin \theta_{k} )$ having a phase shift of $\exp( \jmath \frac{2 \pi}{\lambda} (m-1) d \sin \theta_{k} )$ compared to the reference sensor; $s_{k}$ represents the $k$-th target echo attenuated by the path loss $\beta_{k}$; $w_m$ is the noise contaminating the signal of the $m$-th receives sensor. In the RAQ-ULA system, we assume that all $M$ sensor apertures are identical. Upon using a Gaussian beam of the probe and coupling lasers having the same diameter, we know that each sensor becomes a beam cylinder.

	By further collecting all measurements from the $M$ receiver sensors based on \eqref{eq:NarrowbandSampledOutputPlusNoise}, we arrive at the matrix form 
	\begin{align}
		\label{eq:NarrowbandSampledOutputPlusNoise_MatrixForm}
		\bm{y} = \bm{\varPhi} \bm{A} \left( \bm{\theta} \right) \bm{s} + \bm{w},
	\end{align}
	where $\bm{\varPhi} = \text{diag} \{ \sqrt{\varrho_1} \Phi_1, \sqrt{\varrho_2} \Phi_2, \cdots, \sqrt{\varrho_M} \Phi_M \}$; $\bm{A} \left( \bm{\theta} \right) = [ \bm{a} \left( {\theta_{1}} \right), \bm{a} \left( {\theta_{2}} \right), \cdots, \bm{a} \left( {\theta_{K}} \right) ]$ having its $k$-th vector constituted by the array response vector $\bm{a} \left( {\theta_{k}} \right) = [ 1, \exp( \jmath \frac{2 \pi}{\lambda} d \sin \theta_{k} ),$ $\cdots, \exp( \jmath \frac{2 \pi}{\lambda} (M-1) d \sin \theta_{k} ) ]^{T}$; $\bm{s}$ is the echo vector  and $\bm{w}$ is the noise vector. Specifically, $\bm{w}$ is assumed to obey the complex additive white Gaussian noise (AWGN), namely we have $\bm{w} \sim \mathcal{CN} (\bm{0}, \sigma^2 \bm{I}_{M})$, where $\sigma^2 = \frac{{\cal P}_{\text{QPN}} + {\cal P}_{\text{PSN}} + {\cal P}_{\text{ITN}}}{2}$ denotes the noise power consisting of the quantum projection noise, photon shot noise and intrinsic thermal noise \cite{gong2024RAQRModel_Journal}.

	To gain further insights concerning \eqref{eq:NarrowbandSampledOutputPlusNoise_MatrixForm}, we elaborate on $\bm{\varPhi}$ in more detail. We recall that the LO is assumed to be plane wave, so that the phase at the $m$-th receiver sensor obeys $\theta_{l,m} = \theta_{l,1} + \frac{2 \pi}{\lambda} d (m-1) \sin \vartheta$, where $\theta_{l,1}$ represents the phase at the reference (first) sensor, $\vartheta$ is the DOA of the incident LO signal, and $\lambda$ denotes the wavelength of the LO. As the LO is well-designed, we assume that LO's DOA can be configured in advanced. 
	More particularly, as we have assumed in Section \ref{subsec:QR} that the Rabi frequencies and the frequency detunnings of the probe, coupling, and LO signal are identical for all $M$ receiver sensors, we have ${\cal P}_{m} (\Omega_{l}) \triangleq {\cal P} (\Omega_{l})$, ${\kappa_{m}}({\Omega _l}) \triangleq {\kappa}({\Omega _l})$, and $\varphi_{m} (\Omega_{l}) \triangleq \varphi (\Omega_{l})$ for all $M$ sensors. Let us assume furthermore that the local optical beams are identical for all $M$ receiver sensors in terms of the power and phase, namely we have ${\cal P}_{m}^{(l)} \triangleq {\cal P}_{l}$, $\phi_{m}^{(l)} \triangleq \phi_{l}$. We can reformulate \eqref{eq:GainBCOD}, \eqref{eq:PhaseBCOD} as 
	\begin{align}
		\label{eq:GainBCOD_refoumulate}
		&\varrho_{m} = 4 \alpha_{1}^{2} Z_0 G \mathcal{P}_{l} \mathcal{P} (\Omega_{l}) {\kappa^{2}}({\Omega _l}), \\
		\label{eq:PhaseBCOD_refoumulate}
		&\Phi_{m} = \Phi e^{ - \jmath \frac{2 \pi}{\lambda} d (m-1) \sin \vartheta },  
	\end{align}
	where $\Phi \triangleq \frac{\exp( - \jmath \left[ {\theta_{l,1}} - \varphi ({\Omega _l}) \right] )}{2} + \frac{\exp( - \jmath \left[ {\theta_{l,1}} + \varphi ({\Omega _l}) \right] )}{2}$ is the phase shift of the reference receiver sensor. 
	Based on the above discussions, we obtain $\bm{\varPhi} = \sqrt{\varrho} \Phi \bm{D}$ and reformulate \eqref{eq:NarrowbandSampledOutputPlusNoise_MatrixForm} as 
	\begin{align}
		\label{eq:NarrowbandSampledOutputPlusNoise_MatrixForm_SC}
		\bm{y} = \sqrt{\varrho} \Phi \bm{D} \bm{A} \left( \bm{\theta} \right) \bm{s} + \bm{w}, 
	\end{align}
	where we have $\varrho \triangleq 4 \alpha_{1}^{2} Z_0 G \mathcal{P}_{l} \mathcal{P} (\Omega_{l}) {\kappa^{2}}({\Omega _l})$ and $\bm{D} \triangleq \text{diag}$ $\{ 1, e^{ - \jmath \frac{2 \pi}{\lambda} d \sin \vartheta }, \cdots, e^{ - \jmath \frac{2 \pi}{\lambda} d (M-1) \sin \vartheta } \}$.

	\textbf{Remark 1}: 
	As observed from \eqref{eq:NarrowbandSampledOutputPlusNoise_MatrixForm}, the impinging signals received by different receiver sensors experience different amplitude and phase responses due to $\bm{\varPhi}$. As for \eqref{eq:NarrowbandSampledOutputPlusNoise_MatrixForm_SC}, the amplitude responses become identical and the phase response exists due to $\bm{D}$. We note that both cases introduce a complex-valued \textit{sensor gain mismatch}, i.e., inconsistencies exist in the amplification of signals across the multiple sensors.

	\vspace{-0.3cm}
	\section{RAQ-ULA Based DOA Estimation} 
	\label{sec:DOA}
	
	Upon employing our signal model \eqref{eq:NarrowbandSampledOutputPlusNoise_MatrixForm_SC}, we will estimate the DOAs of all targets with the aid of the pre-designed LO. To this end, we proposed the RAQ-ESPRIT to mitigate the sensor gain mismatch introduced by the LO. To better exhibit RAQ-ESPRIT, we also compare it to the maximum likelihood (ML) estimation and the Cramer-Rao lower bound (CRLB).

	\vspace{-0.3cm}
	\subsection{The Proposed RAQ-ESPRIT} 
	\label{sec:RAQ-ESPRIT}
	We consider two groups of the receives sensors: The first group consists of $M-1$ elements indexed consecutively by $\mathcal{I}_{1} = \{ 1, \cdots, M-1 \}$, while the second group includes another $M-1$ elements indexed consecutively by $\mathcal{I}_{2} = \{ 2, \cdots, M \}$. Therefore, we have their signal models formulated as follows 
	\begin{align}
		\label{eq:SigModel_G1}
		\bm{y}_{1} &= \sqrt{\varrho} \Phi \bm{D}_{1} \bm{A}_{1} \left( \bm{\theta} \right) \bm{s} + \bm{w}_{1}, \\
		\nonumber
		\bm{y}_{2} &= \sqrt{\varrho} \Phi \bm{D}_{2} \bm{A}_{2} \left( \bm{\theta} \right) \bm{s} + \bm{w}_{2} \\
		\label{eq:SigModel_G2}
		&= \sqrt{\varrho} \Phi {e^{ - \jmath \frac{{2\pi }}{\lambda }d\sin \vartheta }} \bm{D}_{1} \bm{A}_{1} \left( \bm{\theta} \right) {\bm{\varTheta}} \bm{s} + \bm{w}_{2}, 
	\end{align}
	where $\bm{D}_{1} = \text{diag} \{ 1, e^{ - \jmath \frac{2 \pi}{\lambda} d \sin \vartheta }, \cdots, e^{ - \jmath \frac{2 \pi}{\lambda} d (M-2) \sin \vartheta } \}$, $\bm{D}_{2} = \text{diag} \{ e^{ - \jmath \frac{2 \pi}{\lambda} d \sin \vartheta }, e^{ - \jmath \frac{2 \pi}{\lambda} 2d \sin \vartheta }, \cdots, e^{ - \jmath \frac{2 \pi}{\lambda} (M-1) d \sin \vartheta } \}$, $\bm{\varTheta} = \text{diag} \{ {e^{\jmath \frac{{2\pi }}{\lambda }d\sin {\theta _1}}}, {e^{\jmath \frac{{2\pi }}{\lambda }d\sin {\theta _2}}}, \cdots, {e^{\jmath \frac{{2\pi }}{\lambda }d\sin {\theta _K}}} \}$, $\bm{A}_{1}, \bm{A}_{2} \in \mathbb{C}^{(M-1) \times K}$ are sub-matrices of $\bm{A}$ with their rows determined by $\mathcal{I}_{1}$ and $\mathcal{I}_{2}$, respectively. \eqref{eq:SigModel_G2} is obtained by exploiting $\bm{D}_{2} = {e^{ - \jmath \frac{{2\pi }}{\lambda }d\sin \vartheta }} \bm{D}_{1}$ and $\bm{A}_{2} \left( \bm{\theta} \right) = \bm{A}_{1} \left( \bm{\theta} \right) {\bm{\varTheta}}$.

	Furthermore, we collect $N$ samples to form the corresponding matrices of $\bm{Y}_{1}$, $\bm{Y}_{2}$, $\bm{S}$, and $\bm{W}$ from their vector counterparts, respectively. Upon stacking the sample matrices $\bm{Y}_{1}$ and $\bm{Y}_{2}$, we arrive at 
	\begin{align}
		\label{eq:SigModel_GM}
		{\bm{Y}} = \sqrt {\varrho} \Phi \bar{\bm{A}} {\bm{S}} + {\bm{W}},  
	\end{align}
	where we define the following matrices 
	\begin{align}
		\nonumber
		{\bm{Y}} \triangleq 
		\begin{bmatrix}
			{{\bm{Y}}_1} \\
			{{\bm{Y}}_2}
		\end{bmatrix}, 
		\bar{\bm{A}} \triangleq 
		\begin{bmatrix}
			{{\bm{D}}_1} {{\bm{A}}_1} \left( \bm{\theta} \right) \\
			{e^{ - \jmath \frac{{2\pi}}{\lambda}d\sin \vartheta}} {{\bm{D}}_1} {{\bm{A}}_1} \left( \bm{\theta} \right) {\bm{\Theta}}
		\end{bmatrix}, 
		{\bm{W}} \triangleq 
		\begin{bmatrix}
			{{\bm{W}}_1} \\
			{{\bm{W}}_2}
		\end{bmatrix}. 
	\end{align}
	\vspace{-1em} 
	\begin{theorem}
		\label{theorem1}
		Consider the noiseless counterpart of \eqref{eq:SigModel_GM} expressed in the form of ${\bm{Y}} = \sqrt {\varrho} \Phi \bar{\bm{A}} {\bm{S}}$ and represent the singular value decomposition of ${\bm{Y}}$ in the form of ${\bm{Y}} = {\bm{U}} {\bm{\varSigma}} {\bm{V}}^{\text{H}}$. We prove that $\bm{U}$ and $\bar{\bm{A}}$ have the same column space, namely $span \left\{ \bm{U} \right\} = span \left\{ \bar{\bm{A}} \right\}$. Furthermore, we have a relationship of ${\bm{U}} = \bar{\bm{A}} {\bm{T}}$ for some invertible matrix ${\bm{T}} \in {\mathbb{C}^{K \times K}}$. 
	\end{theorem}
	\vspace{-0.3em}
	\begin{IEEEproof}
		Since ${{\bm{A}}_1}\left( {\bm{\theta }} \right) \in \mathbb{C}^{(M-1) \times K}$ has a full column rank of $K$ and ${{\bm{D}}_1} \in \mathbb{C}^{(M-1) \times (M-1)}$ is diagonal with non-zero elements, we thus know that ${{\bm{D}}_1}{{\bm{A}}_1}\left( {\bm{\theta }} \right)$ is also a full column rank matrix and we have $rank\left\{ {{\bm{D}}_1}{{\bm{A}}_1}\left( {\bm{\theta }} \right) \right\} = rank\left\{ {{\bm{A}}_1}\left( {\bm{\theta }} \right) \right\} = K$. Similarly, we can prove that ${e^{ - \jmath \frac{{2\pi}}{\lambda}d\sin \vartheta}} {{\bm{D}}_1} {{\bm{A}}_1} \left( \bm{\theta} \right) {\bm{\Theta}}$ has full column rank and $rank\left\{ {e^{ - \jmath \frac{{2\pi}}{\lambda}d\sin \vartheta}} {{\bm{D}}_1} {{\bm{A}}_1} \left( \bm{\theta} \right) {\bm{\Theta}} \right\} = K$. Therefore, we know that $rank\left\{ \bar{\bm{A}} \right\} = K$. 
		Furthermore, $\bm{S}$ is a full row rank matrix for $N > K$, namely $rank\left\{ \bm{S} \right\} = K$. Based on the above discussion, it is obvious that the noiseless $\bm{Y}$ has a rank of $K$ and $span \{ \bm{Y} \} = span \{ \bar{\bm{A}} \}$. Upon exploiting the relationship of $span \{ \bm{U} \} = span \{ \bm{Y} \}$, we obtain that $span \left\{ \bm{U} \right\} = span \left\{ \bar{\bm{A}} \right\}$. 
		
		More particularly, as revealed by linear algebra, a matrix can be expressed as a linear transformation of another matrix if they have the same column space. Specifically, we have some invertible matrix $\bm{T} \in {\mathbb{C}^{K \times K}}$ that serves as the above-mentioned linear transformation to enable the relationship of $\bm{U} = \bar{\bm{A}} \bm{T}$, where the inversibility of $\bm{T}$ is guaranteed by the full column rank property of $\bm{U}$ and $\bar{\bm{A}}$. 
	\end{IEEEproof}

	It is noted that ${\bm{U}}$ can be divided into two sub-matrices ${\bm{U}_{1}} \in {\mathbb{C}^{(M-1) \times K}}$ and ${\bm{U}_{2}} \in {\mathbb{C}^{(M-1) \times K}}$ corresponding to the two groups of the receiver sensors, respectively. Based on \textbf{Theorem \ref{theorem1}},  we have the following relationship
	\begin{align}
		\label{eq:relationship1}
		{{\bm{U}}_1} &= {{\bm{D}}_1} {{\bm{A}}_1} \left( \bm{\theta} \right) {\bm{T}}, \\
		\label{eq:relationship2}
		{{\bm{U}}_2} &= {e^{ - \jmath \frac{{2\pi}}{\lambda}d\sin \vartheta}} {{\bm{D}}_1} {{\bm{A}}_1} \left( \bm{\theta} \right) {\bm{\Theta}} {\bm{T}}.
	\end{align}
	Upon denoting the Moore-Penrose pseudoinverse of ${{\bm{U}}_1}$ by ${{\bm{U}}_{1}^{\dagger}}$, we can directly arrive at 
	\begin{align}
		\label{eq:relationship3}
		{{\bm{U}}_{1}^{\dagger}} {{\bm{U}}_2} =  {{\bm{T}}^{-1}} \left( e^{ - \jmath \frac{{2\pi }}{\lambda} d \sin \vartheta } {\bm{\varTheta }} \right) {\bm{T}}, 
	\end{align}
	where we have exploited that  ${{\bm{D}}_1^{\rm{H}}{{\bm{D}}_1}} = {\bm{I}_{(M-1) \times (M-1)}}$. 
	It is explicit that \eqref{eq:relationship3} can be interpreted as the eigen-decomposition of ${{\bm{U}}_{1}^{\dagger}} {{\bm{U}}_2}$, when $\bm{T}$ is unitary, $\bm{T}^{-1} = \bm{Q}$ and $\bm{T} = \bm{Q}^{H}$, where $\bm{Q}$ represents the matrix consisting of eigenvectors of ${{\bm{U}}_{1}^{\dagger}} {{\bm{U}}_2}$. 
	Furthermore, let us denote the eigenvalues of the matrix ${{\bm{U}}_{1}^{\dagger}} {{\bm{U}}_2}$ by $\{ \sigma_{1}, \sigma_{2}, \cdots, \sigma_{K} \}$, where we have $\sigma_k = e^{ - \jmath \frac{{2\pi }}{\lambda} d \sin \vartheta } [\bm{\varTheta}]_{k,k}$. Consequently, the DOA of the $k$-th target is estimated as 
	\begin{align}
		\label{eq:doa_estimate}
		{\theta _k} = \arcsin \left[ \frac{\lambda }{2\pi d} \angle { \left( e^{ \jmath \frac{{2\pi }}{\lambda} d \sin \vartheta } \sigma_k \right)} \right], 
	\end{align}
	where $\angle (\cdot)$ represents the angle of a complex value.

	\textbf{Remark 2}: 
	Upon directly employing the classical ESPRIT, one can obtain the estimated DOA as ${\theta _k} = \arcsin \left( \frac{\lambda }{2\pi d} \angle {\sigma_k} \right)$. By contrast, the DOA obtained by the proposed RAQ-ESPRIT compensates the DOA estimated by the classical ESPRIT by applying the multiplicative correction term $e^{ \jmath \frac{{2\pi }}{\lambda} d \sin \vartheta }$ related to the LO, as seen from \eqref{eq:doa_estimate}. This compensation mitigates the sensor gain mismatch degradation and facilitates an enhanced estimation performance. Additionally, the computational complexity of the RAQ-ESPRIT is comparable to that of the traditional ESPRIT because they process the same data size, and perform a similar subspace decomposition. 

	\vspace{-0.8em}
	\subsection{Maximum Likelihood and Cramer-Rao Lower Bound}
	\label{sec:mle_crb}
	
	Let us assume that the signals $\bm{s}$ are deterministic. Then the DOAs of $K$ targets estimated by the ML can be obtained by maximizing the problem $\bm{\theta}_{\rm{RAQ-ML}} = \arg \max_{\bm{\theta}} \; \Tr ( \bm{P} \hat{\bm{R}}_{y} )$, where $\bm{P} = \bm{D} \bm{A} \left( \bm{\theta} \right) [ \bm{A}^{H} \left( \bm{\theta} \right) \bm{A} \left( \bm{\theta} \right) ]^{-1} \bm{A}^{H} \left( \bm{\theta} \right) \bm{D}^{H}$ represents the projection matrix and $\hat{\bm{R}}_{y} = \frac{1}{N} \bm{Y} \bm{Y}^{H}$ is the sampled covariance matrix. The asymptotic ML estimation error for classical antenna arrays has been shown in \cite[eq. 2.15]{stoica1990performance}. Following a similar derivation process, we can obtain the multi-target estimation error of the ML for RAQ-ULAs as 
	\begin{align}
		\nonumber
		&\hspace{-0.6em} \varepsilon_{\rm{RAQ-ML}} = \varpi \times \\
		\label{eq:errorMLE}
		&\hspace{-1.4em} \Tr \left(
		\left[ \mathscr{R} \left\{\bm{H} \odot \hat{\bm{R}}_{s}^{T} \right\} \right]^{-2} 
		\mathscr{R} \left\{\bm{H} \odot \left( \hat{\bm{R}}_{s} \bm{W}(\varpi) \hat{\bm{R}}_{s} \right)^{T} \right\}
		\right), 
	\end{align}
	where $\varpi = \frac{\sigma^2}{2 N \varrho \Phi^{H} \Phi}$, $\bm{H} = \dot{\bm{A}}^{H} \left( \bm{\theta} \right) \bm{D}^{H} \left( \bm{I} - \bm{P} \right) \bm{D} \dot{\bm{A}} \left( \bm{\theta} \right)$, $\dot{\bm{A}} \left( \bm{\theta} \right) = \left[ \frac{ \partial \bm{a}(\theta_{1}) }{ \partial \theta_{1} }, \frac{ \partial \bm{a}(\theta_{2}) }{ \partial \theta_{2} }, \cdots, \frac{ \partial \bm{a}(\theta_{K}) }{ \partial \theta_{K} } \right]$, $\hat{\bm{R}}_{s} = \frac{1}{N} \bm{S} \bm{S}^{H}$, and $\bm{W}(\varpi) = \hat{\bm{R}}_{s}^{-1} + 2N \varpi \hat{\bm{R}}_{s}^{-1} \left[ \bm{A}^{H} \left( \bm{\theta} \right) \bm{A} \left( \bm{\theta} \right) \right]^{-1} \hat{\bm{R}}_{s}^{-1}$. 
	Furthermore, the CRLB reflects the lower bound of the multi-target estimation error. Following a similar derivation process as in \cite[eq. 2.11]{stoica1990performance}, we express the CRLB for RAQ-ULA as 
	\begin{align}
		\label{eq:CRB}
		\varepsilon_{\rm{RAQ-CRLB}} &= \varpi \Tr \left(
		\left[ \mathscr{R} \left\{ \bm{H} \odot \hat{\bm{R}}_{s}^{T} \right\} \right]^{-1} 
		\right). 
	\end{align}
	
	As seen from equations \textcolor{red}{(44), (45)} of \cite{gong2024RAQRModel_Journal}, the received signal-to-noise-ratio (SNR) of RAQRs has been studied in the photon shot limit (PSL) and the standard quantum limit (SQL), respectively. Based on these results, we can reformulate $\varpi$ as 
	\begin{align}
		\label{eq:varpi}
		\varpi = \left\{ 
		\begin{aligned}
			&\frac{B}{ 2N {{\cal P}}({\Omega_l}) \kappa^2({\Omega_l}) \cos^2 {\varphi ({\Omega _l})} }, &\rm{PSL}, \\
			&\frac{1}{{4{Z_0}N}} {\left( \frac{\hbar}{\mu_{34}} \right)^2} \left( \frac{{\varGamma_2}}{{{\bar N}_0} V} \right) B, &\rm{SQL},
		\end{aligned} \right. 
	\end{align}
	where $B$ represents the bandwidth of the targets' signals, $\bar{N}_0 = \varUpsilon N_0$ is the effective atomic density, $\varUpsilon$ is the atomic excitation fraction, $\varGamma_{2}$ is the total dephasing rate, and $V$ is the cylindrical volume of a single receiver sensor. Therefore, we can substitute the two values of $\varpi$ back into \eqref{eq:errorMLE} and \eqref{eq:CRB} to obtain the estimation error of the ML and the CRLB in the PSL and SQL regimes, respectively. 
	More particularly, in the PSL regime, both $\varepsilon_{\rm{RAQ-ML}}$ and $\varepsilon_{\rm{RAQ-CRLB}}$ can be minimized by ensuring that $\cos^2 {\varphi_{m} ({\Omega _l})} = \cos^2 {\varphi ({\Omega _l})} = 1$. This can be realized to adjust the phase of the local optical beam ${\phi_{m}^{(l)}}$ to retain ${\varphi_{m} ({\Omega _l})} = 0$ based on \eqref{eq:varphi_2}.

	\section{Simulation Results}
	\label{sec:Simulations}
	
	In this section we present simulations quantifying its DOA estimation error versus (vs.) diverse parameters.
	
	\vspace{-0.2cm}
	\subsection{Simulation Configurations}
	
	In the following simulations, we use a vapour cell having a length of $\ell = 10$ cm filled with Cesium (Cs) atoms at an atomic density of $N_0 = 4.89 \times 10^{10}$ $\text{cm}^{-3}$ and a total atomic excitation fraction of $1\%$. The inter-sensor spacing is half-wavelength of the targets' signals, where their carrier frequency and bandwidth are $6.9458$ GHz and $100$ kHz, respectively.  
	The four-level transition system of \ref{fig:RAQMIMOScheme} is 6S\textsubscript{\scalebox{0.8}{1/2}} $\rightarrow$ 6P\textsubscript{\scalebox{0.8}{3/2}} $\rightarrow$ 47D\textsubscript{\scalebox{0.8}{5/2}} $\rightarrow$ 48P\textsubscript{\scalebox{0.8}{3/2}}. Following TABLE \textcolor{red}{I} of \cite{gong2024RAQRModel_Journal}, the dipole moment and decay rate of each energy level, the total dephasing rate, the wavelength/power/radius of the probe and coupling beams, the amplitude of the LO, and the parameters of the PDA are provided. The phase shift of the LO is set to $\pi/3$. As obtained in \cite{gong2024RAQRModel_Journal}, the laser detunings are optimized as $\Delta_{p,c,l} = \{-0.9133, 1.8090, -0.0075\}$ MHz. 
	For comparisons, the classical RF receiver array is configured as the base station at the frequency range of 5G FR1 n104, where the parameters are specified by 3GPP \cite{3GPP_IMT} and provided in TABLE \textcolor{red}{I} of \cite{gong2024RAQRModel_Journal}.

	Furthermore, we assume that multiple targets are randomly distributed within a circular area that has a radius of $500$ meters and the RAQ-ULA is $1500$-meter away from the center of the circular. The impinging DOAs are randomly generated within $-90^{\circ}$ to $90^{\circ}$. Furthermore, we consider line-of-sight (LoS) propagation between the RAQ-ULA and the targets. The path loss imposed to the target echoes is computed by $K_0 + 10v \log \frac{u}{u_0}$ with $K_0 = -30$, $v = 2$, $u_0 = 1$ meter, and $u$ represents the distance between the RAQ-ULA and the targets. We characterize the mean squared error (MSE), namely $\varepsilon = \mathbb{E}\{\| \bm{\theta} - \hat{\bm{\theta}} \|^2\}$, in our simulations, where the results are averaged over $500$ realizations. Unless otherwise stated, we set $M=10$, $K=5$, $N=50$, and $23$ dBm reflected power.

	\begin{figure*}[t!]
		\centering
		\subfloat{
			\includegraphics[width=0.326\textwidth]{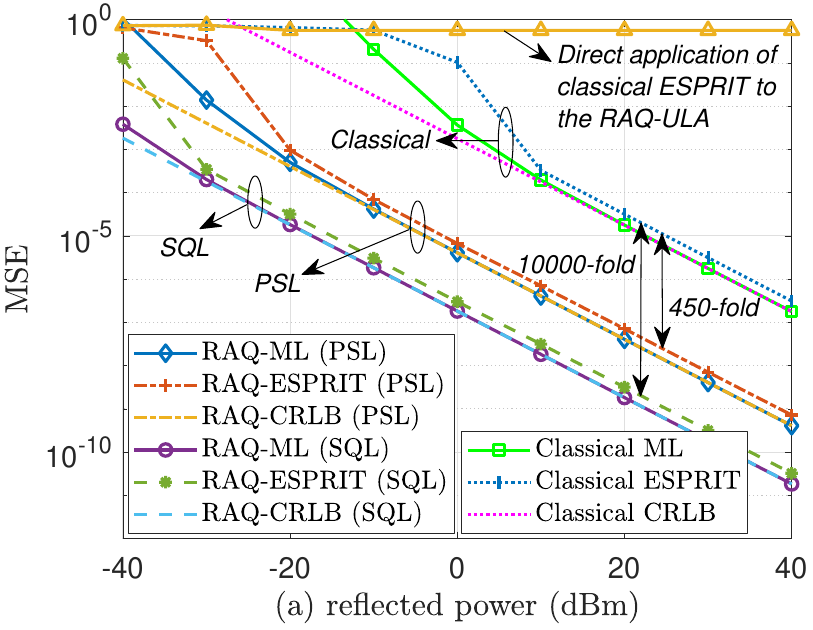}} 
		\subfloat{
			\includegraphics[width=0.326\textwidth]{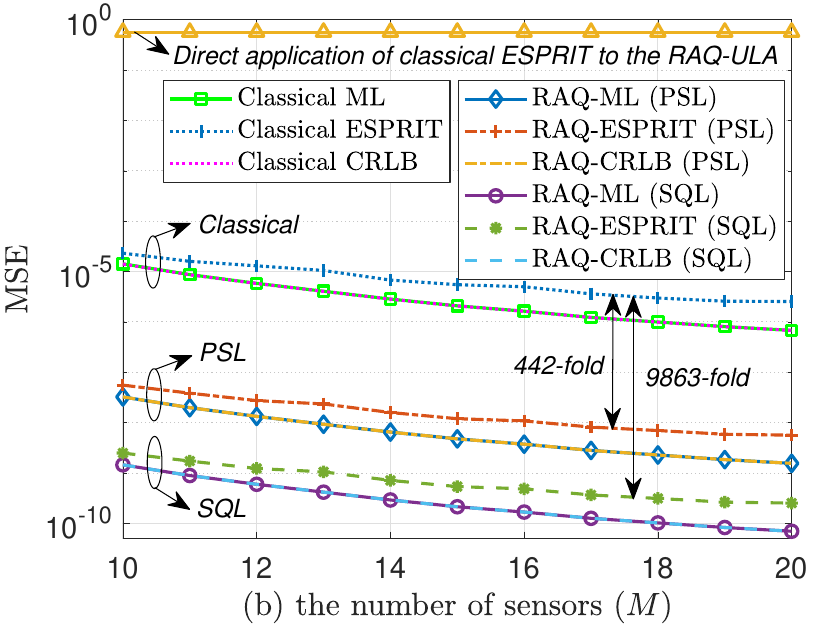}} 
		\subfloat{
			\includegraphics[width=0.326\textwidth]{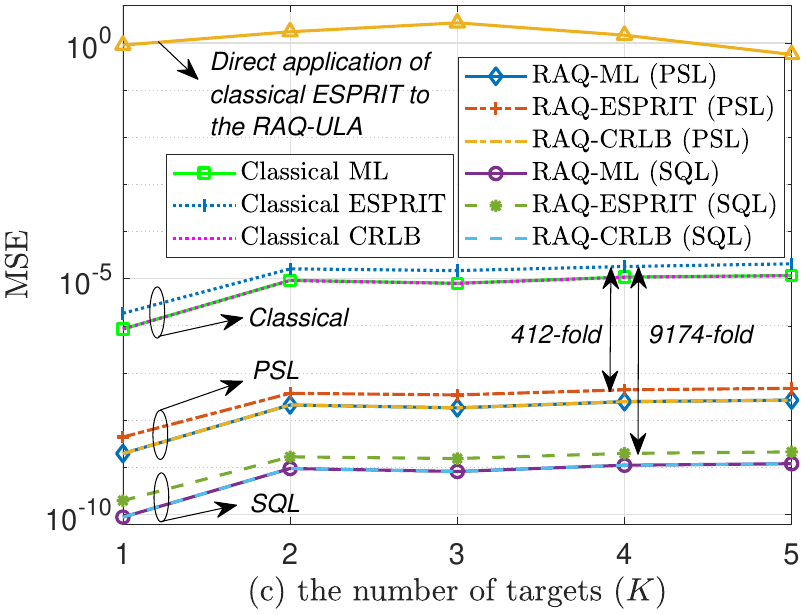}} \\
		\subfloat{
			\includegraphics[width=0.326\textwidth]{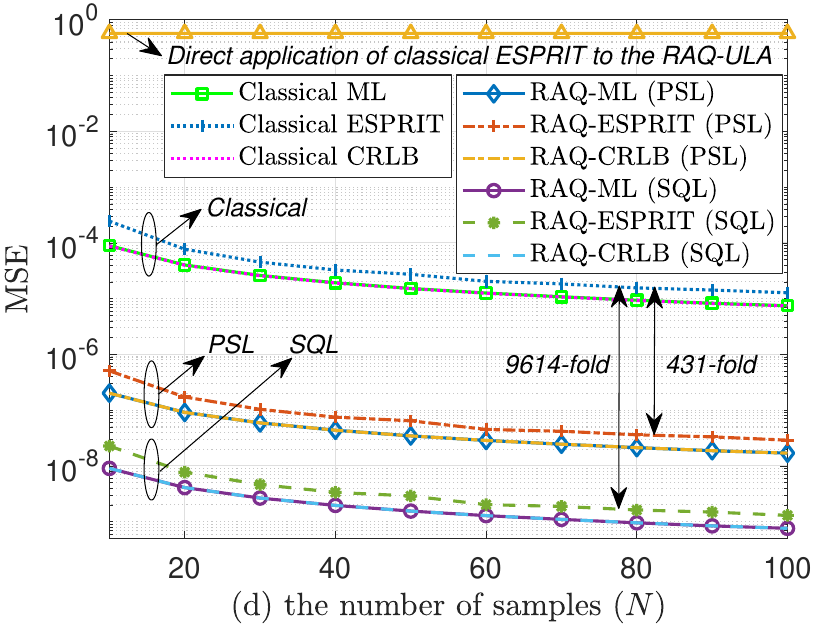}} 
		\subfloat{
			\includegraphics[width=0.326\textwidth]{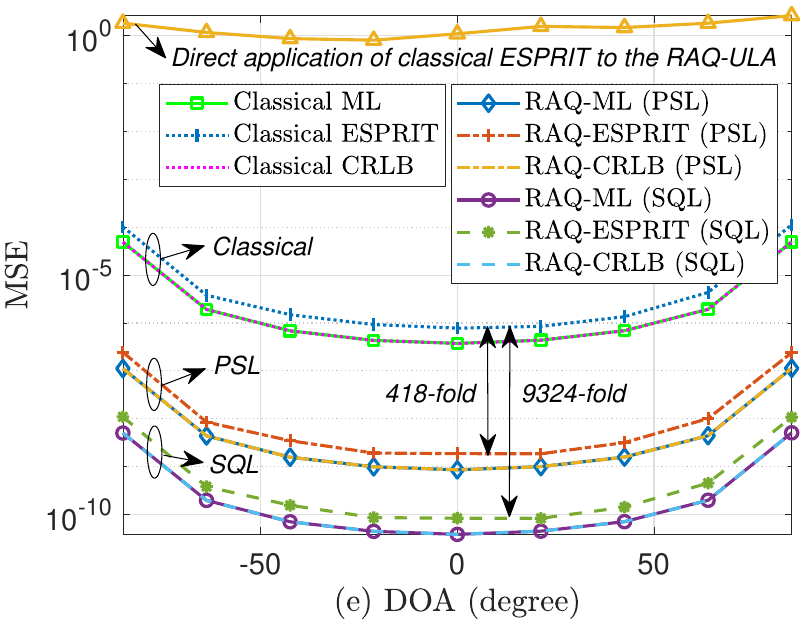}}
		\subfloat{
			\includegraphics[width=0.326\textwidth]{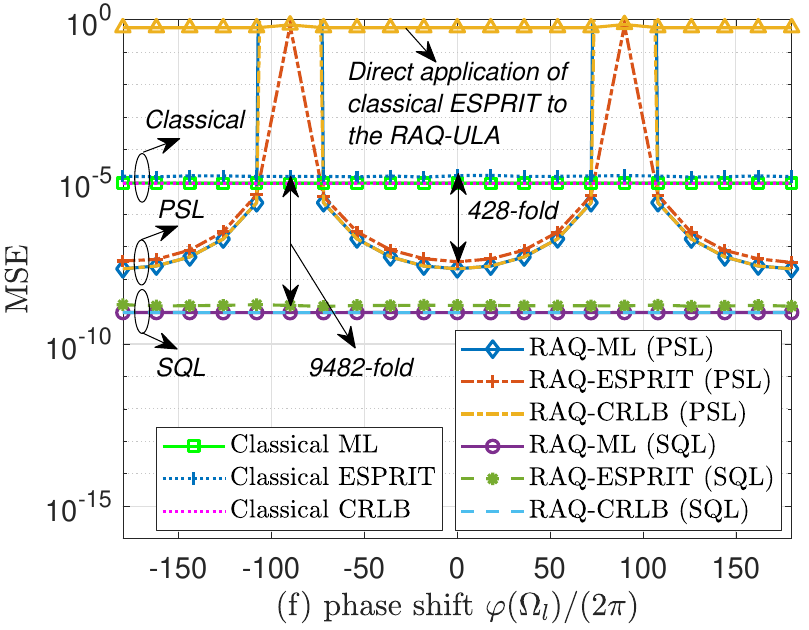}}
		\vspace{-0.3em}
		\caption{The MSE vs. (a) the reflected power of $s_k$, (b) the number of sensors ($M$), (c) the number of targets ($K$), (d) the number of samples ($N$), (e) the range of DOA, and (f) the phase shift ${\varphi ({\Omega _l})}$.}
		\vspace{-1.2em}
		\label{fig:TransmitPower&NumberOfSensors&targets&samples}
	\end{figure*}

	\vspace{-0.2cm}
	\subsection{Simulation Results}
	
	We present the simulation results in Fig. \ref{fig:TransmitPower&NumberOfSensors&targets&samples} for the RAQ-ESPRIT, RAQ-ML, RAQ-CRLB, and for the classical counterparts for antenna-based ULAs. We also plot the curve of directly applying the classical ESPRIT to the RAQ-ULA for showing its infeasibility. 
	Upon varying the power reflected from the targets, we observe from Fig. \ref{fig:TransmitPower&NumberOfSensors&targets&samples}(a) that the RAQ-ESPRIT exhibits a significant reduction in the MSE, which can be on the order of $450$-fold and $10000$-fold in the PSL and SQL regimes, respectively. We note that the reflected power affects the received SNR, which is different for the RAQ-ULA and for the classical RF receiver due to the different noise floor experienced and the receiver gain. 
	Then, we determine the number of sensors $M$ in Fig. \ref{fig:TransmitPower&NumberOfSensors&targets&samples}(b). As $M$ increases, all the MSE curves decrease, but the RAQ-ESPRIT significantly outperforms its classical counterpart, where the MSE is reduced on the order of $442$-fold and $9863$-fold in the PSL and SQL regimes, respectively. 
	Next, we present the MSE vs. the number of targets $K$ in Fig. \ref{fig:TransmitPower&NumberOfSensors&targets&samples}(c) and vs. the number of samples $N$ in Fig. \ref{fig:TransmitPower&NumberOfSensors&targets&samples}(d), respectively. We observe that all curves increase as $K$ grows, whereas the MSE decreses as $N$ increases. For Fig. \ref{fig:TransmitPower&NumberOfSensors&targets&samples}(c)(d), RAQ-ESPRIT has a much lower MSE, both showing a significant MSE reduction of $> 400$-fold and $> 9000$-fold in the PSL and SQL regimes, respectively. 
	Furthermore, the MSE vs. the DOA $\theta_{k}$ is portrayed in Fig. \ref{fig:TransmitPower&NumberOfSensors&targets&samples}(e). It is observed that the RAQ-ESPRIT more significantly reduces the MSE than its classical counterpart over the whole DOA range, exhibiting $418$-fold and $9324$-fold in the PSL and SQL regimes, respectively. 
	We note that Fig. \ref{fig:TransmitPower&NumberOfSensors&targets&samples}(a)-(e) are obtained when $\cos^2 {\varphi ({\Omega _l})} = 1$. To characterize the influence of ${\varphi ({\Omega _l})}$, we present the MSE vs. ${\varphi ({\Omega _l})} / (2 \pi)$ in Fig. \ref{fig:TransmitPower&NumberOfSensors&targets&samples}(f). It is observed that the MSE in the PSL regime becomes unbounded when ${\varphi ({\Omega _l})} / (2 \pi) = 90^{\circ}$, while it is minimally achieved when ${\varphi ({\Omega _l})} / (2 \pi) = 0^{\circ}$, $\pm 180^{\circ}$, as indicated by \eqref{eq:varpi}.

		\section{Conclusions}
		\label{sec:Conclusions}
		
		In this article, we have conceived RAQRs for the classical DOA estimation problem. To this end, we have designed a RAQ-ULA architecture for detecting multiple targets and constructed its equivalent baseband signal model accordingly. Our signal model is consistent with the actual implementation of a RAQR, paving the way for future RAQR aided wireless sensing designs. Based on our model, we have also proposed a RAQ-ESPRIT method for DOA estimation to mitigate the sensor gain mismatch problem. Lastly, we have performed simulations for demonstrating the superiority of our scheme.

\balance 
\bibliographystyle{IEEEtran}
\bibliography{IEEEabrv,references} 


\end{document}